\documentclass[preprint2]{aastex62}

\newcommand\aastex{AAS\TeX}

\usepackage{epstopdf}

\received{July 1, 2019}
\revised{July 1, 2019}
\accepted{\today}

\submitjournal{ApJ}

\shorttitle{\aastex\ Energy partition in circular-ribbon flares}
\shortauthors{Zhang et al.}

\begin{document}

\title{Energy partition in two M-class circular-ribbon flares}

\correspondingauthor{Q. M. Zhang}
\email{zhangqm@pmo.ac.cn}

\author[0000-0003-4078-2265]{Q. M. Zhang}
\affil{Key Laboratory for Dark Matter and Space Science, Purple Mountain Observatory, CAS, Nanjing 210034, People's Republic of China}
\affil{State Key Laboratory of Space Weather, Chinese Academy of Sciences, Beijing 100190, People's Republic of China}

\author{J. X. Cheng}
\affil{Key Laboratory of Planetary Sciences, Shanghai Astronomical Observatory, Shanghai 200030, People's Republic of China}

\author[0000-0003-4655-6939]{L. Feng}
\affil{Key Laboratory for Dark Matter and Space Science, Purple Mountain Observatory, CAS, Nanjing 210034, People's Republic of China}
\affil{School of Astronomy and Space Science, University of Science and Technology of China, Hefei, Anhui 230026, People's Republic of China}

\author[0000-0002-4241-9921]{Y. Su}
\affil{Key Laboratory for Dark Matter and Space Science, Purple Mountain Observatory, CAS, Nanjing 210034, People's Republic of China}
\affil{School of Astronomy and Space Science, University of Science and Technology of China, Hefei, Anhui 230026, People's Republic of China}

\author[0000-0002-3032-6066]{L. Lu}
\affil{Key Laboratory for Dark Matter and Space Science, Purple Mountain Observatory, CAS, Nanjing 210034, People's Republic of China}

\author{Y. Huang}
\affil{Key Laboratory for Dark Matter and Space Science, Purple Mountain Observatory, CAS, Nanjing 210034, People's Republic of China}
\affil{School of Astronomy and Space Science, University of Science and Technology of China, Hefei, Anhui 230026, People's Republic of China}

\author[0000-0002-4538-9350]{D. Li}
\affil{Key Laboratory for Dark Matter and Space Science, Purple Mountain Observatory, CAS, Nanjing 210034, People's Republic of China}
\affil{State Key Laboratory of Space Weather, Chinese Academy of Sciences, Beijing 100190, People's Republic of China}

\author{T. H. Zhou}
\affil{Key Laboratory for Dark Matter and Space Science, Purple Mountain Observatory, CAS, Nanjing 210034, People's Republic of China}

\author{J. L. Chen}
\affil{Key Laboratory for Dark Matter and Space Science, Purple Mountain Observatory, CAS, Nanjing 210034, People's Republic of China}
\affil{School of Astronomy and Space Science, University of Science and Technology of China, Hefei, Anhui 230026, People's Republic of China}

\begin{abstract}
In this paper, we investigate the energy partition of two homologous M1.1 circular-ribbon flares (CRFs) in active region (AR) 12434. 
They were observed by \textit{SDO}, \textit{GOES}, and \textit{RHESSI} on 2015 October 15 and 16, respectively.
The peak thermal energy, nonthermal energy of flare-accelerated electrons, total radiative loss of hot plasma, and radiant energies in 1$-$8 {\AA} and 1$-$70 {\AA} of the flares are calculated.
The two flares have similar energetics. The peak thermal energies are (1.94$\pm$0.13)$\times$10$^{30}$ erg.
The nonthermal energies in flare-accelerated electrons are (3.9$\pm$0.7)$\times$10$^{30}$ erg.
The radiative outputs of the flare loops in 1$-$70 {\AA}, which are $\sim$200 times greater than the outputs in 1$-$8 {\AA}, account for $\sim$62.5\% of the peak thermal energies.
The radiative losses of SXR-emitting plasma are one order of magnitude lower than the peak thermal energies.
Therefore, the total heating requirements of flare loops including radiative loss are (2.1$\pm$0.1)$\times$10$^{30}$ erg, which could sufficiently be supplied by nonthermal electrons.
\end{abstract}


\section{Introduction} \label{s:intro}
Solar flares and coronal mass ejections (CMEs) are the most spectacular and energetic activities in our solar system, which have potential impact on space weather \citep{sch06,chen11,fle11,hol11}.
The free magnetic energy is accumulated before flares via various mechanisms \citep{wie14}, such as flux emergence \citep{sun12}, twisting \citep{li17a,xu17}, 
and/or shearing motions in the photosphere \citep{su07}. The stored energy is impulsively released via magnetic reconnection \citep{pri02,su13,xue16,li17b} and converted to thermal energy 
of the directly heated plasma, kinetic energy of the bulk flow, and nonthermal energies of the accelerated electrons as well as ions \citep{asch02,mann06}.
The total energy content of flares ranges from $\sim$10$^{24}$ erg for the smallest ones (nanoflares) to $\sim$10$^{33}$ erg for the largest ones (X-class flares).
The nonthermal electrons spiral down along the newly reconnected magnetic field lines and precipitate in the dense chromosphere, resulting in significant chromospheric evaporation 
and/or condensation \citep{bro71,fis85,ning09,li15,benz17,tian18}. In this way, the energy in nonthermal electrons is converted through Coulomb collisions into energy in the thermal plasma.

A fraction of energetic electrons may escape into the interplanetary space along the open field and generate type III radio bursts \citep{mann99,benz08}.
The radiations in hard X-ray (HXR), soft X-ray (SXR), ultraviolet (UV), extreme-ultraviolet (EUV), white light (WL), infrared, and radio wavelengths increase dramatically and reach an apex 
during the impulsive phase that spans a few to tens of minutes \citep[][see Fig. 2 for an illustration]{benz08}. Afterwards, the fluxes of radiation decline with time and the post-flare loops cool down 
via heat conduction and radiative loss \citep{car95}. Flares are classified into confined and eruptive types according to their association with CMEs \citep{wang07,sun15,th15}, 
although the visibility of CMEs depends greatly on flare intensity \citep{yas05}.

Although a general framework of flares has been established, the energy partition is still controversial in that the energy release, transport, and conversion processes are 
very complicated and strongly coupled. In fact, the situation varies from case to case. Hence, statistical studies are more appropriate to address this issue.
After a detailed investigation of 24 flares observed by \textit{GOES} and 
the \textit{Ramaty Hight Energy Solar Spectroscopic Imager} \citep[\textit{RHESSI};][]{lin02}, \citet{war16a} found that a relatively cooler plasma component (10$-$25 MK)
is produced by chromospheric evaporation, while a hotter component ($\geq$25 MK) is produced by direct heating in the corona. Therefore, they concluded that 
electron beam heating at the chromosphere is insufficient to account for the heating of the hot thermal plasma and supplying the bolometric radiation.
Conductive loss ($\propto T^{7/2}$) provides an efficient way of transporting energy from the corona to the lower atmosphere so that the energy is quickly radiated away
in UV, EUV, and WL \citep{war16b}.

Using coordinated observations from multiple space-borne instruments, \citet{ems04} evaluated the energetics of two flare/CME events, including the energy contents of
CMEs, thermal plasma of the flares, nonthermal electrons, ions, and solar energetic particles. It is concluded that, with large uncertainties, CMEs contain a dominant component 
of the released free energy. After refining the flare energy estimates, \citet{ems05} came to a conclusion that flare and CME energies ($\sim$10$^{32}$ erg) are comparable, 
which was further substantiated by the analysis of a different flare/CME event by \citet{feng13}.
\citet{ems12} carried out a comprehensive investigation of the energetics of 38 large eruptive flares during 2002 February and 2006 December. It is revealed that the energy 
content in flare-accelerated electrons and ions is sufficient to supply the bolometric radiant energy. 

In a big project, \citet{asch14,asch15,asch16,asch17} studied the global energetics of flares and CMEs observed by the Atmospheric Imaging Assembly \citep[AIA;][]{lem12} 
on board the \textit{Solar Dynamics Observatory} (\textit{SDO}) during the first 3.5 yr of its mission. The multiwavelength observations of AIA facilitate the calculation of 
thermal energy using the differential emission measure (DEM) distribution function. It is shown that the multithermal DEM function yields a considerably higher (multi-)thermal energy
than an isothermal energy of the flares based on the same AIA data. Moreover, the nonthermal energy exceeds the thermal energy in 85\% of the whole events.

In the context of standard flare model, the flare ribbons show diverse motions \citep{fle11,hol16}.
In most cases, the ribbons separate as magnetic reconnection proceeds \citep{qiu02,ji04}. Apart from separation, the brightening may propagate along the ribbons \citep{qiu17}.
For three-dimensional magnetic reconnection within the thin quasi-separatrix layers \citep{dem96}, flare ribbons propagate along the intersection of QSLs with the chromosphere \citep{aul07}.
Circular-ribbon flares (CRFs) are a special kind of flares that consist of a central short ribbon and a surrounding ribbon with 
a circular or quasi-circular shape \citep[e.g.,][]{mas09,reid12,jia13,jos15,kum15,yang15,hao17,her17,song18,chen19,li19,zqm19a}. Most of them are confined flares without CMEs.
The magnetic topology of CRFs is usually composed of a null point, a spine, and a dome-like fan surface \citep{zqm12,zqm15,mas17,li18}. 
In 2015 October, a series of homologous, short-lived CRFs occurred in active region (AR) 12434. For the first time, \citet{zqm16a} studied explosive chromospheric evaporation 
in a C4.2 CRF that occurred on October 16. Besides, \citet{zqm16b} reported periodic chromospheric condensation in a homologous C3.1 CRF in the same AR.
So far, the energy partition in CRFs has not been explored. 

In this paper, we study the energy partition in two homologous M1.1 CRFs on October 15 and 16, respectively.
In Section~\ref{s:data}, we describe the observations and data analysis. The morphological evolution of the flares and accompanying type III radio bursts are presented in Section~\ref{s:crf}. 
The calculations of different energy contents are elucidated in Section~\ref{s:eng}. We compare our findings with previous works in Section~\ref{s:disc}
and give a brief summary in Section~\ref{s:sum}.

\section{Observations and data analysis} \label{s:data}
The M1.1 CRFs taking place in AR 12434 were observed by \textit{SDO}, \textit{RHESSI}, \textit{WIND}, and the Nobeyama Radioheliograph \citep[NoRH;][]{naka94}.
AIA takes full-disk images in two UV (1600 and 1700 {\AA}) and seven EUV (94, 131, 171, 193, 211, 304, and 335 {\AA}) wavelengths. The photospheric line-of-sight (LOS) magnetograms 
were provided by the Helioseismic and Magnetic Imager \citep[HMI;][]{sch12} on board \textit{SDO}. The AIA and HMI level\_1 data were calibrated using the standard Solar Software (SSW)
programs \textit{aia\_prep.pro} and \textit{hmi\_prep.pro}, respectively. The fluxes of the flares in 1$-$70 {\AA} were recorded by the Extreme Ultraviolet Variability Experiment \citep[EVE;][]{wood12} 
on board \textit{SDO}. SXR fluxes of the flares in 0.5$-$4 {\AA} and 1$-$8 {\AA} were recorded by \textit{GOES}. The isothermal temperature ($T_{e}$) and emission measure (EM) of the 
SXR-emitting plasma can be derived from the ratio of \textit{GOES} fluxes \citep{wh05}. 

To obtain the HXR light curves, images, and spectra evolution, we use observations from \textit{RHESSI}.
HXR images near the flare peak times were reconstructed using the CLEAN method with an integration time of 4 s \citep{hur02}.
The pulse pileup correction, energy gain correction, and isotropic albedo photo correction are carried out to obtain the background-subtracted spectra. In order to derive the properties of 
accelerated electrons, we fit the HXR spectra by the combination of an isothermal component determined by the isothermal temperature and EM and a nonthermal component created by
the thick-target bremsstrahlung of energetic electrons with a low-energy cutoff \citep{bro71,war16a}. The spectra fitting is conducted using the OSPEX software built in SSW.

As a ground-based radio telescope at the Nobeyama Radio Observatory, NoRH observes the Sun at frequencies of 17 and 34 GHz with spatial resolutions of 10$\arcsec$ and 5$\arcsec$, respectively. 
In addition, the type III radio bursts related to the flares were recorded in the radio dynamic spectra by \textit{WIND}/WAVES \citep{bou95}.
WAVES consists of two detectors: RAD1 (0.02$-$1.04 MHz) and RAD2 (1.075$-$13.825 MHz). The observational parameters are summarized in Table~\ref{tab:para}.

\begin{deluxetable}{cccc}
\tablecaption{Description of the observational parameters \label{tab:para}}
\tablecolumns{4}
\tablenum{1}
\tablewidth{0pt}
\tablehead{
\colhead{Instrument} &
\colhead{$\lambda$} &
\colhead{Cadence} & 
\colhead{Pixel Size} \\
\colhead{} & 
\colhead{({\AA})} &
\colhead{(s)} & 
\colhead{(\arcsec)}
}
\startdata
\textit{SDO}/AIA  & 94$-$335 &  12  & 0.6 \\
\textit{SDO}/AIA  & 1600       &  24   & 0.6 \\
\textit{SDO}/HMI & 6173      &  45    & 0.6 \\
\textit{SDO}/EVE & 1$-$70 &  0.25 & \nodata \\
\textit{GOES}     & 0.5$-$4 &  2.05 & \nodata \\
\textit{GOES}     & 1$-$8    &  2.05 & \nodata \\
\textit{RHESSI}   & 3$-$50 keV &  4.0 & 2.0 \\
NoRH                 & 17 GHz &  1     & 5.0 \\
\textit{WIND}/WAVES & 0.02$-$13.825 MHz &  60  & \nodata \\
\enddata
\end{deluxetable}

\section{Circular-ribbon flares and type III radio bursts} \label{s:crf}
\begin{figure*}
\plotone{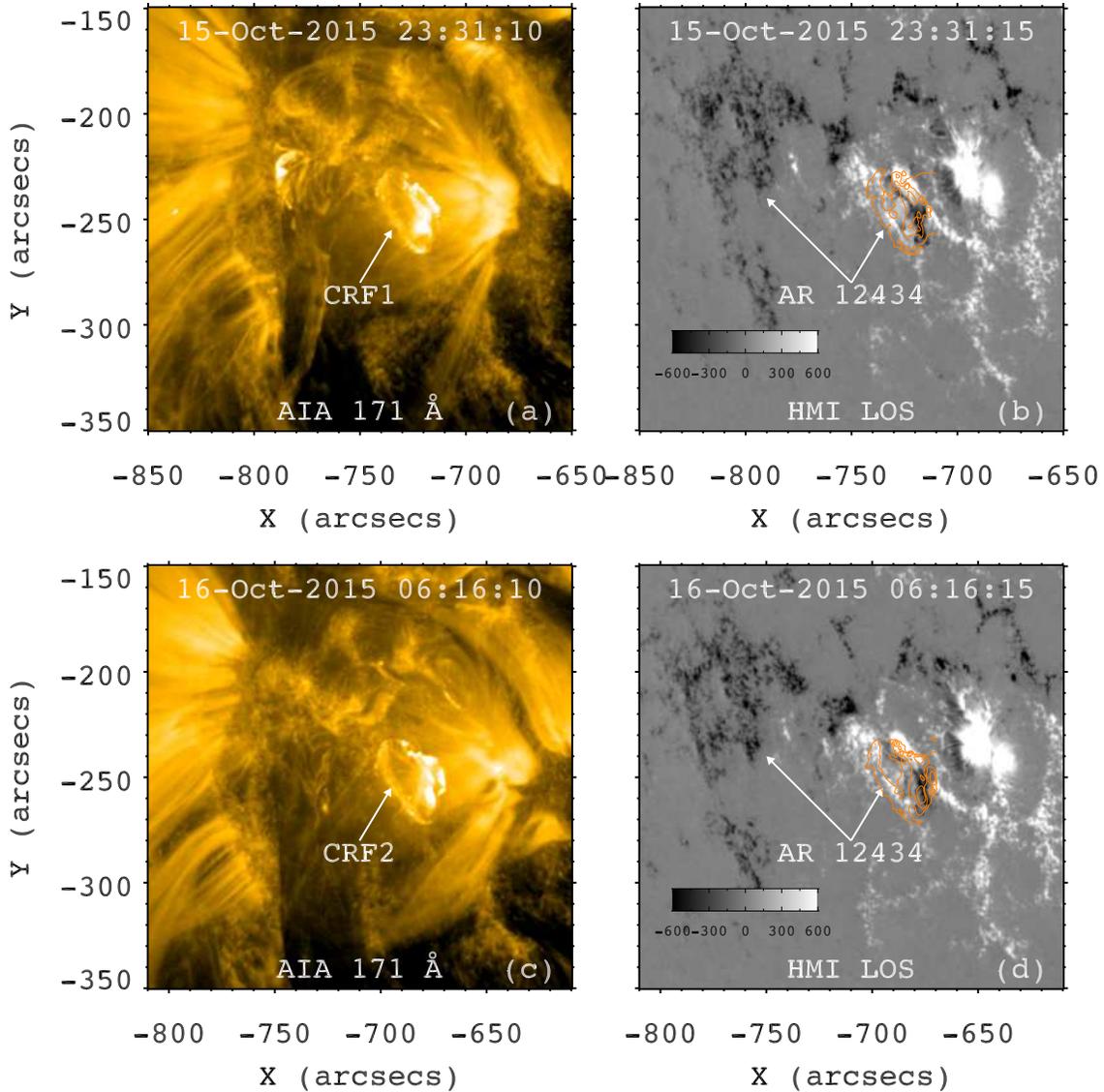}
\caption{Top panels: AIA 171 {\AA} image and HMI LOS magnetogram around 23:31:10 UT on 2015 October 15.
CRF1 and AR 12434 are indicated by the arrows. 
Bottom panels: AIA 171 {\AA} image and HMI LOS magnetogram around 06:16:10 UT on 2015 October 16.
CRF2 and AR 12434 are indicated by the arrows. Intensity contours of the EUV images are superposed on the magnetograms with orange lines.
(An animation of this figure is available.)
\label{fig1}}
\end{figure*}

In Figure~\ref{fig1}, AIA 171 {\AA} images of the flares (CRF1 and CRF2) are displayed in the left panels (see also the online animation).  
Like the C-class flares reported in \citet{zqm16a,zqm16b}, CRF1 and CRF2 feature a short inner ribbon and an 
elliptical outer ribbon. The length scale ($\sim$40$\arcsec$) of these compact flares is comparable to that of coronal bright points \citep{zqm12}. 
The corresponding HMI LOS magnetograms are displayed in the right panels of Figure~\ref{fig1}.
The inner and outer ribbons correspond to negative and positive polarities in the photosphere, respectively.

\begin{figure*}
\plotone{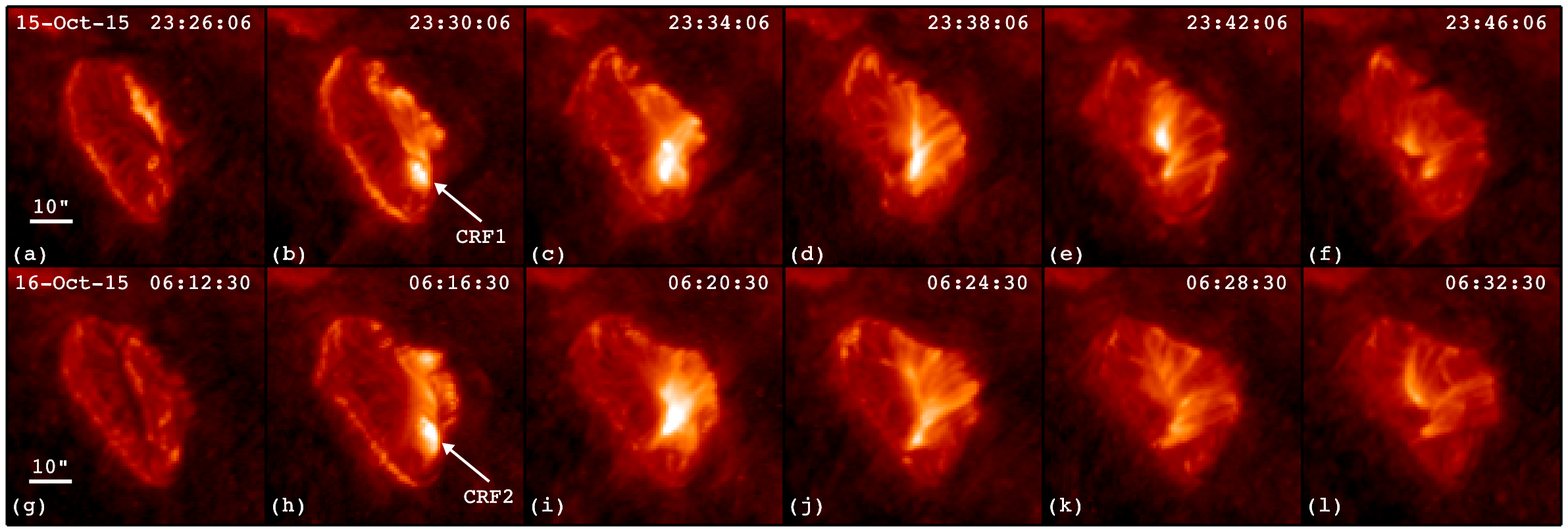}
\caption{Snapshots of the AIA 304 {\AA} images with a field of view of 60$\arcsec$$\times$60$\arcsec$ during CRF1 (top panels) and CRF2 (bottom panels).
(An animation of this figure is available.)
\label{fig2}}
\end{figure*}

Figure~\ref{fig2} shows twelve snapshots of the AIA 304 {\AA} images (see also the online animation).
The two flares have some kind of similarity in morphological evolution. The inner ribbon and northwest part of outer ribbon brightened first, which was followed by brightening 
of the southeast part of outer ribbon (see panels (b) and (h)). Afterwards, the brightness of flare ribbons declined with time. Since the ribbon separation in confined flares is more or less static
\citep{hint18}, the areas of CRFs did not change remarkably as expected.

\begin{figure}
\plotone{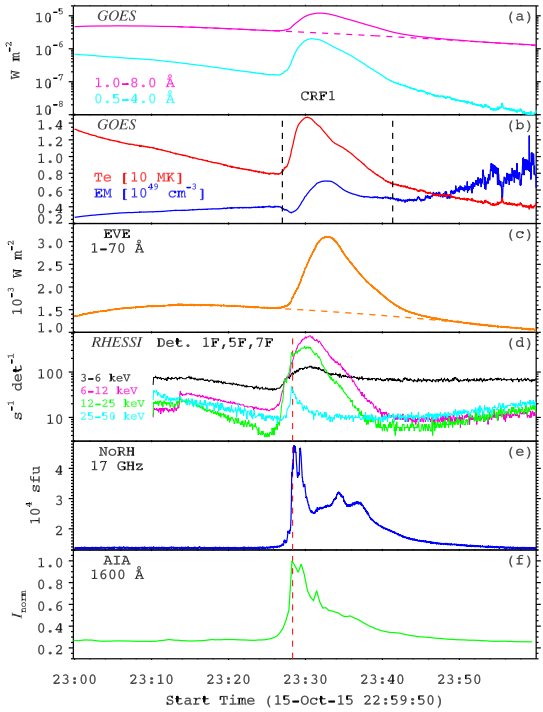}
\caption{(a) SXR light curves of CRF1 in 0.5$-$4 {\AA} and 1$-$8 {\AA}. The magenta dashed line represents the background intensity.
(b) Time evolutions of temperature ($T_e$) and emission measure (EM) derived from \textit{GOES} observations.
The black dashed lines represent the lower and upper limits of integrals.
(c) Light curve of CRF1 in 1$-$70 {\AA}. The orange dashed line represents the background intensity.
(d) HXR light curves of CRF1 at different energy bands.
(e) Radio light curve of CRF1 in 17 GHz.
(f) UV light curve of CRF1 in 1600 {\AA}.
The red dashed line in panels (d)-(f) denote the peak time at 23:28:20 UT.
\label{fig3}}
\end{figure}

In Figure~\ref{fig3}, the SXR light curves of CRF1 are plotted in panel (a). The SXR flux in 1$-$8 {\AA} increases rapidly from $\sim$23:27 UT until the peak value at $\sim$23:31 UT, 
before a gradual decline until $\sim$23:50 UT. Therefore, the lifetime of CRF1 is $\sim$23 minutes. 
The dashed line signifies the background flux, which will be subtracted to get a net SXR irradiance (see Figure~\ref{fig7}(a)).
Figure~\ref{fig3}(b) shows the time evolutions of the isothermal temperature ($T_e$) and EM of the SXR-emitting plasma, which will be used to calculate the total radiated energy 
from the SXR-emitting plasma. The temperature reaches a peak value of 14.7 MK at 23:30:15 UT. The EM reaches a peak value of 0.7$\times$10$^{49}$ cm$^{-3}$ at 23:32:26 UT.
The light curve in 1$-$70 {\AA} is displayed in Figure~\ref{fig3}(c), with a peak value at 23:32:54 UT. The dashed line represents the background flux, which will be subtracted to
get a net irradiance (see Figure~\ref{fig7}(c)). The corrected count rates at different energy bands are plotted with different colors in Figure~\ref{fig3}(d). 
The peak time of nonthermal flux at 25$-$50 keV at $\sim$23:28:20 UT is indicated by the red dashed line. 

In Figure~\ref{fig5}, the full-disk radio image in 17 GHz at 23:28:30 UT is demonstrated in panel (a). A closeup of CRF1 is shown in panel (b). Due to the lower resolution
of NoRH compared with AIA, the flare is a single, bright source. Time evolution of the integral flux of CRF1 is plotted in Figure~\ref{fig3}(e). The peak time of radio flux coincides with 
that of HXR flux at 25$-$50 keV, confirming the nonthermal nature of emissions from the flare-accelerated, high-energy electrons.
Figure~\ref{fig3}(f) shows the time evolution of normalized integral intensity of CRF1 in 1600 {\AA}.

\begin{figure}
\plotone{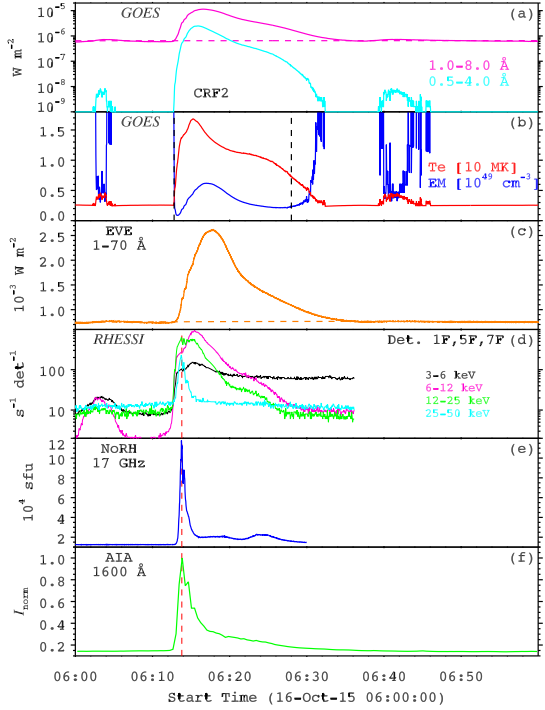}
\caption{Same as Figure~\ref{fig3}, but for CRF2. The red dashed line in panels (d)-(f) denote the peak time at 06:13:48 UT.
\label{fig4}}
\end{figure}

\begin{figure}
\plotone{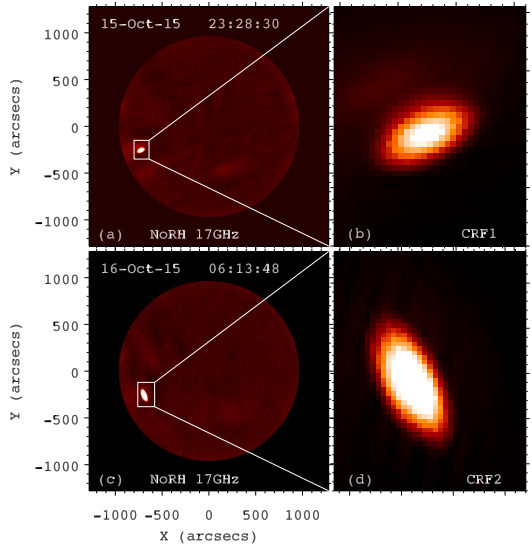}
\caption{Left panels: Full-disk radio images in 17 GHz on October 15 and 16.
Right panels: Closeups of CRF1 and CRF2.
\label{fig5}}
\end{figure}

Like in Figure~\ref{fig3}, the light curves of CRF2 in various wavelengths are drawn in Figure~\ref{fig4}. The SXR flux in 1$-$8 {\AA} starts to increase at $\sim$06:11 UT and 
reaches the apex at $\sim$06:16 UT, followed by a gradual decline until $\sim$06:35 UT. Therefore, the lifetime of CRF2 is $\sim$24 minutes.
The maximal isothermal temperature (16.8 MK) of CRF2 is slightly higher than that of CRF1,
while the maximal EM of CRF2 is slightly lower than that of CRF1. In Figure~\ref{fig5}, the full-disk radio image in 17 GHz at 06:13:48 UT on October 16 is displayed in panel (c), 
and a closeup of CRF2 is displayed in panel (d). Time evolution of the integral radio flux of CRF2 is plotted in Figure~\ref{fig4}(e). 
Likewise, the time evolution of normalized integral intensity of CRF2 in 1600 {\AA} is shown in Figure~\ref{fig4}(f). 
The simultaneous peak time at 06:13:48 UT is indicated by the red dashed line.

\begin{figure}
\plotone{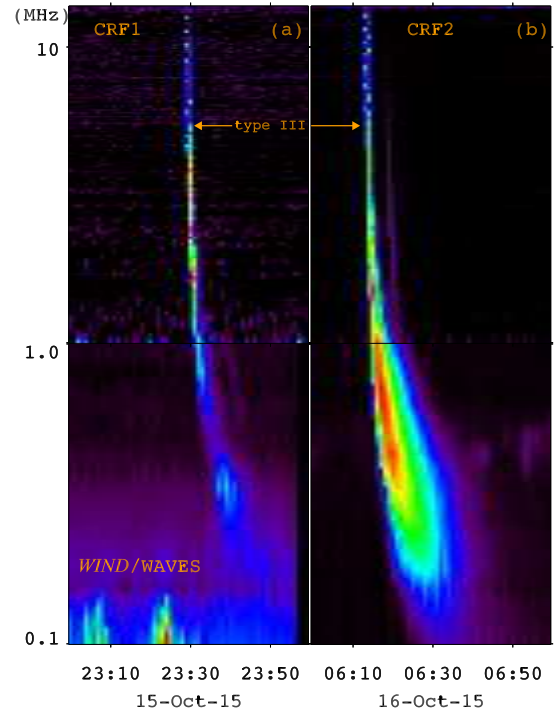}
\caption{Radio dynamic spectra recorded by \textit{WIND}/WAVES during CRF1 and CRF2.
The arrows point to the type III radio bursts associated with the flares.
\label{fig6}}
\end{figure}

Like the C3.1 CRF reported by \citet{zqm16b}, both CRF1 and CRF2 were associated with type III radio bursts, which are pointed by the arrows in the radio dynamic spectra recorded by
\textit{WIND}/WAVES (see Figure~\ref{fig6}). The starting times of radio bursts were consistent with the peak times in 17 GHz, implying a common origin of the radio emissions. 
The flare-accelerated nonthermal electrons streaming down into the chromosphere create strong emission in 17 GHz via gyro-synchrotron radiation mechanism, while those escaping from the
corona along open field lines create type III bursts via plasma radiation mechanism.

\section{Energy partition} \label{s:eng}

\subsection{Radiated energy in \textit{GOES} 1$-$8 {\AA}} \label{s:sxr18}
In this Section, we focus on the different energy contents of the confined flares. First of all, we calculate the radiated energy in \textit{GOES} 1$-$8 {\AA}.
As mentioned in \citet{ems12} and \citet{feng13}, the contribution of background radiation should be removed. 
In the original 1$-$8 {\AA} light curves, a few data points before $t_{sta}$ and after $t_{end}$ in Table~\ref{tab:eng} are extracted and used for a quadratic fitting.
The background light curves are obtained based on the results of fitting. The background-subtracted light curves in 1$-$8 {\AA}
($f_{1-8}$) are plotted in the top panels of Figure~\ref{fig7}. The radiated energy is calculated by integrating the background-subtracted light curve,
\begin{equation} \label{eqn1}
  U_{1-8}=2\pi d^2\int_{t_1}^{t_2}f_{1-8}(t)dt,
\end{equation}
where $d\approx1.496\times10^8$ km (1 AU) represents the distance between the Sun and the Earth.
The lower limit ($t_1$) of integral is defined to be the start time of a flare, which is listed in the fourth column of Table~\ref{tab:eng}.
The upper limit ($t_2$) of integral is defined to be the time when $f_{1-8}$ drops to 10\% of the maximal value.
The time of $t_2$ is 23:41:20 UT for CRF1 and 06:28:00 UT for CRF2 (see also top panels of Figure~\ref{fig7}).
The calculated radiated energies, being 5.50$\times$10$^{27}$ erg for CRF1 and 6.24$\times$10$^{27}$ erg for CRF2, are listed in the seventh column of Table~\ref{tab:eng}. 
A bar chart of the energy contents is shown in Figure~\ref{fig12}.

The definition of $t_2$ may influence the value of $U_{1-8}$. For example, if $t_2$ is taken to be the time when $f_{1-8}$ drops to 5\% of the maximal value, 
then $U_{1-8}$ increases by a factor of 1.8\% for CRF1 and 2\% for CRF2, respectively.

\subsection{Radiated energy in EVE 1$-$70 {\AA}} \label{s:xuv170}
\begin{figure}
\plotone{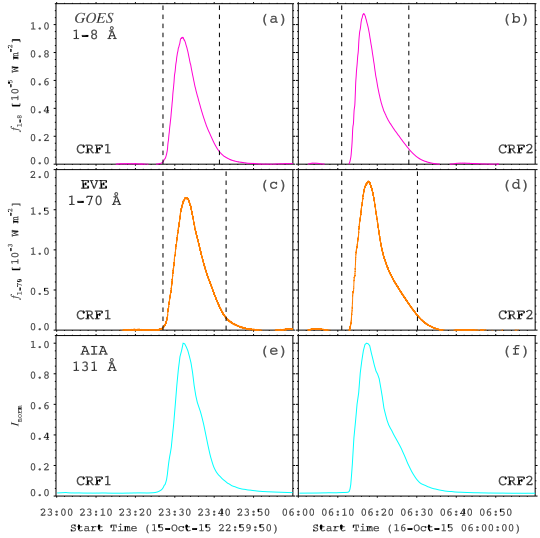}
\caption{(a)-(d) Background-subtracted light curves in 1$-$8 {\AA} and 1$-$70 {\AA} of the two flares.
The dashed lines denote the lower and upper time limits of integrals.
(e)-(f) Light curves of the flares in 131 {\AA}. 
\label{fig7}}
\end{figure}

Like in 1$-$8 {\AA}, we extract a few data points from the original 1$-$70 {\AA} light curves before performing a quadratic fitting and calculating the background light curves.
The background-subtracted light curves in 1$-$70 {\AA} ($f_{1-70}$) are plotted in the middle panels of Figure~\ref{fig7}. 
The radiated energies are calculated by integrating the background-subtracted light curves,
\begin{equation} \label{eqn2}
  U_{1-70}=2\pi d^2\int_{t_1}^{t_2}f_{1-70}(t)dt.
\end{equation}
The lower and upper limits of integral have similar definitions as those in 1$-$8 {\AA}.
The times of $t_1$ are listed in the fourth column of Table~\ref{tab:eng}.
The time of $t_2$ is 23:43:00 UT for CRF1 and 06:30:10 UT for CRF2 (see also middle panels of Figure~\ref{fig7}).
The calculated radiated energies, being 1.12$\times$10$^{30}$ erg for CRF1 and 1.28$\times$10$^{30}$ erg for CRF2, are listed in the eighth column of Table~\ref{tab:eng}. 
It is noted that the total radiated energies in 1$-$70 {\AA} are $\sim$200 times larger than those in 1$-$8 {\AA}.

Similarly, the definition of $t_2$ may influence the value of $U_{1-70}$. If $t_2$ is taken to be the time when $f_{1-70}$ drops to 5\% of the maximal value, 
then $U_{1-70}$ increases by a factor of 1.8\% for CRF1 and 2.3\% for CRF2, respectively.

Unfortunately, there were no observations of flux in 70$-$370 {\AA} from \textit{SDO}/EVE during the flares, so that precise calculation of the radiated energy in 70$-$370 {\AA} 
is unavailable. However, assuming that the ratio ($\sim$30) of radiation in 1$-$70 {\AA} to that in 70$-$370 {\AA} for X-class flares \citep{feng13} is applicable to CRFs, 
the radiated energies in 70$-$370 {\AA} of CRF1 and CRF2 are roughly estimated to be 3.73$\times$10$^{28}$ erg and 4.26$\times$10$^{28}$ erg, respectively. 
The total radiated energies in 1$-$370 {\AA} of CRF1 and CRF2 amount to 1.16$\times$10$^{30}$ erg and 1.32$\times$10$^{30}$ erg accordingly.
\citet{wood04} investigated the contribution of solar UV variation to the total solar irradiance (TSI) of the X17 flare on 2003 October 28. It is found that the combined contribution of 
radiation from XUV (0$-$270 {\AA}) and EUV (270$-$1200 {\AA}) to TSI is $\sim$20\%. That is to say, TSI is a factor of $\sim$5 larger than the radiation in 0$-$1200 {\AA}, 
in which the radiation in 1$-$370 {\AA} plays a predominant role. Based on this assumption, we can deduce the TSI for CRF1 and CRF2, being 5.8$\times$10$^{30}$ erg
and 6.6$\times$10$^{30}$ erg (see Figure~\ref{fig12}).

\subsection{Radiated energy from the SXR-emitting plasma} \label{s:radloss}
\begin{figure}
\plotone{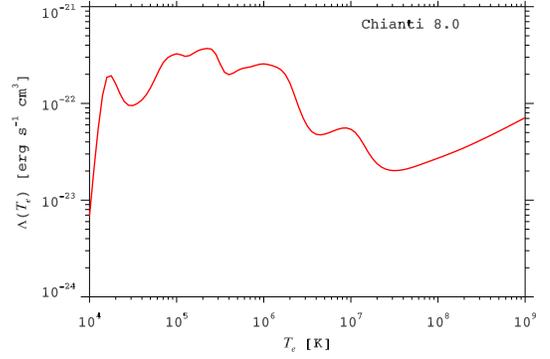}
\caption{Radiative loss rate $\Lambda(T_{e})$ as a function of temperature ($T_e$) 
obtained from CHIANTI  8.0 database.
\label{fig8}}
\end{figure}

The total optically thin radiative loss from the SXR-emitting plasma can be obtained based on the observed EM and $\Lambda(T_{e})$, 
\begin{equation} \label{eqn3}
  T_{rad}=\int_{t_1}^{t_2}\mathrm{EM}(t)\times\Lambda(T_{e}(t))dt,
\end{equation}
where $\Lambda(T_{e})$ denotes the radiative loss rate \citep{cox69}. EM and $T_{e}$ are derived from \textit{GOES} observations (see Figure~\ref{fig3}(b) and Figure~\ref{fig4}(b)).
The lower limit ($t_1$) and upper limit ($t_2$) of the integral are the same as those in Equation~\ref{eqn1}.
For CRF2, the profiles of $T_e$ and EM are abnormal before 06:12:48 UT. Therefore, the time of $t_1$ for CRF2 is taken to be 06:12:48 UT.
In Figure~\ref{fig8}, the profile of $\Lambda(T_{e})$ obtained from CHIANTI  8.0 database assuming coronal abundances is adopted for calculation \citep{del15}.
The total radiative loss from the hot plasma of CRF1 and CRF2, being 2.26$\times$10$^{29}$ erg and 1.41$\times$10$^{29}$ erg, are listed in the ninth column of Table~\ref{tab:eng}.
The radiative losses of the flares are $\sim$40 and $\sim$20 times larger than the radiations in 1$-$8 {\AA} for CRF1 and CRF2, which is consistent with previous results that
radiative output in 1$-$8 {\AA} is one order of magnitude lower than the radiative loss \citep{feng13}. The values of $T_{rad}$ increase by a factor of $\sim$12\% when $t_2$ changes
like in Section~\ref{s:sxr18}.

\subsection{Peak thermal energy of the SXR-emitting plasma} \label{s:th}
The thermal energy of the SXR-emitting plasma can be expressed as
\begin{equation} \label{eqn4}
  E_{th}=3n_{e}k_{B}T_{e}fV=3k_{B}T_{e}\sqrt{\mathrm{EM}\times fV},
\end{equation}
where $n_{e}=\sqrt{\mathrm{EM}/V}$ is the electron number density, $k_B=1.38\times10^{-16}$ erg K$^{-1}$ is the Boltzmann constant, $f\approx1.0$ is the volumetric filling factor, 
and $V$ is the volume of the SXR-emitting plasma \citep{ems12,feng13}. 
Assuming that $V$ is an invariable, then $E_{th}$ reaches the peak value when $T_{e}\sqrt{\mathrm{EM}}$ is maximal (see Figure~\ref{fig3}(b) and Figure~\ref{fig4}(b)). 

\begin{figure*}
\plotone{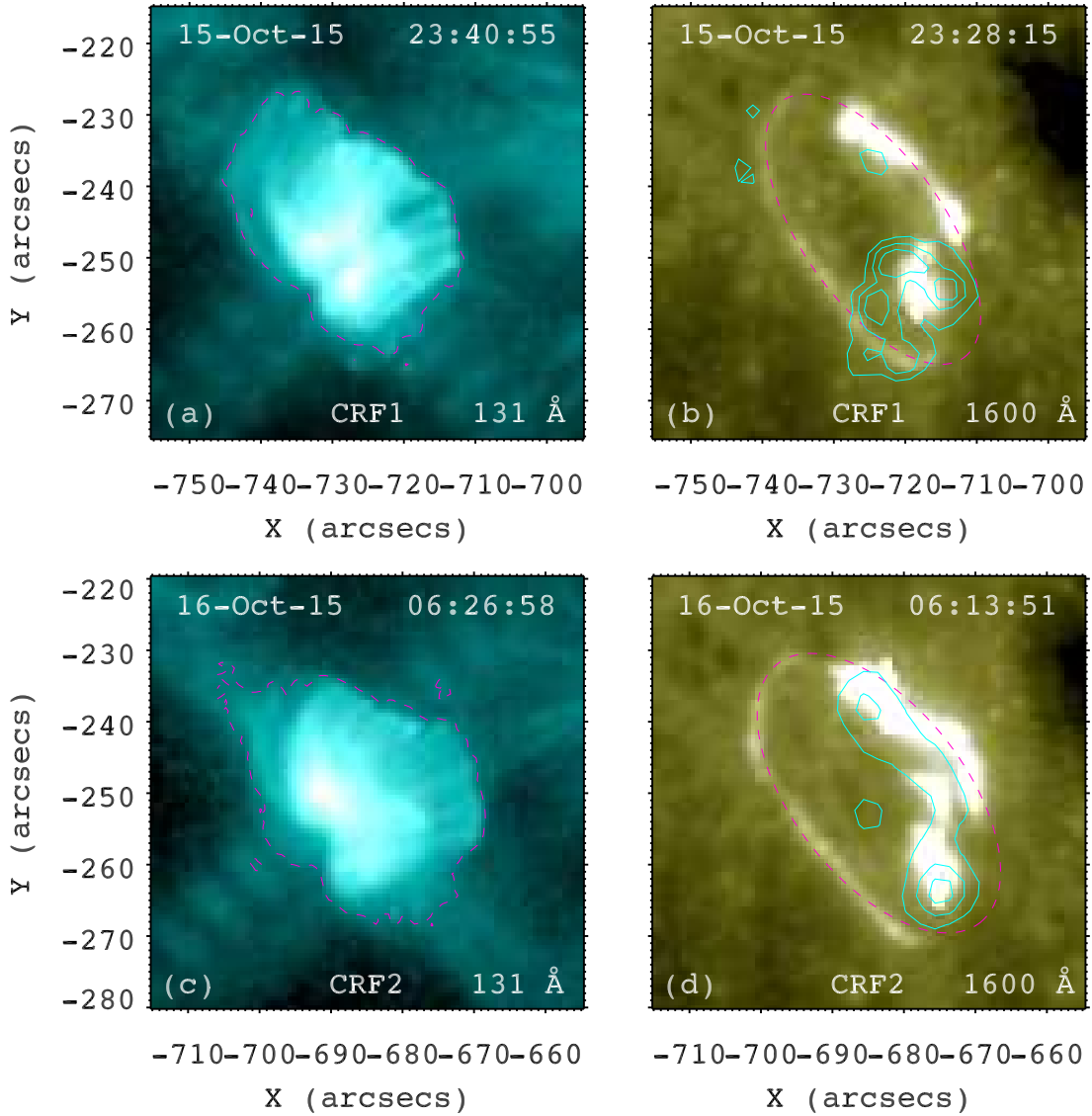}
\caption{Left panels: AIA 131 {\AA} images of the flares $\sim$10 minutes after the peak times. 
Intensity contours of the images are overlaid with magenta dashed lines. The flare area in 131 {\AA} is calculated by summing up the pixels inside the contours.
Right panels: AIA 1600 {\AA} images of the flares at their peak times. Intensity contours of the HXR images at 12$-$25 keV are overlaid with cyan solid lines.
Ellipse fittings of the outer ribbons are overlaid with magenta dashed lines. The flare area in 1600 {\AA} is represented by the area of ellipses.
(An animation of this figure is available.)
\label{fig9}}
\end{figure*}

To get a better estimation of $V$, we use two independent methods. As pointed out by \citet{od10}, the dominant contribution of flare spectrum for AIA 131 {\AA} channel comes from 
the Fe\,{\sc xxi} line with formation temperature of $\log T\approx7.05$. In the bottom panels of Figure~\ref{fig7}, light curves of the flares in 131 {\AA} are plotted with cyan lines.
It is evident that the 131 {\AA} irradiance is well correlated with $f_{1-8}$. Therefore, the hot plasma observed in 131 {\AA} during a flare is a good 
proxy of the SXR-emitting plasma \citep{fle13}. In Figure~\ref{fig9}, the left panels demonstrate the 131 {\AA} images 10 minutes after the peak times ($t_{peak}$ in Table~\ref{tab:eng}), 
since the images at the peak times are severely saturated. The total areas ($A_{131}$) of the flares can be obtained by setting a threshold of intensity, which is $\sim$5 times larger
than the average intensity of a nearby quiet region.
Considering that the flares were close to the limb, the projection effect is corrected by multiplying a factor of $(\cos \mu)^{-1}$, where $\mu$ denotes the longitude of flare core.
The corresponding volume $V_{131}$ is defined as $A_{131}^{3/2}$. The values of $A_{131}$ and $V_{131}$ are listed in the second and third columns of Table~\ref{tab:vol}.
If the threshold intensity is set to be $\sim$6 times larger than the quiet region intensity, the areas in 131 {\AA} drop by a factor of 6\%$-$8\% and the volumes decrease by a factor of
10\%$-$12\%, accordingly.

The second approach to calculating the flare area is based on the fact that the outer ribbons of CRFs mapping the footprints of the post-flare loops in the chromosphere hardly expand with time.
In Figure~\ref{fig9}, the right panels demonstrate the 1600 {\AA} images at their peak times when the ribbons are the most noticeable (see Figure~\ref{fig3}(f) and Figure~\ref{fig4}(f)). 
Since the outer ribbons are not fully closed, we perform an ellipse fitting of the outer ribbons (magenta dashed lines). Hence, the flare areas in 1600 {\AA} ($A_{1600}$) are represented 
by the areas of ellipses after correcting the projection effect. The corresponding volume $V_{1600}$ is defined as $A_{1600}^{3/2}$. 
The values of $A_{1600}$ and $V_{1600}$ are listed in the fourth and fifth columns of Table~\ref{tab:vol}. It is obvious that both area and volume in 131 {\AA} and 1600 {\AA} are close to each other, 
suggesting that the two methods are reasonable. By adopting the mean values of $\bar{V}$ in the last column of Table~\ref{tab:vol} and peak EM during the flares, the mean electron 
number density ($\bar{n}_{e}$) of the hot plasma is estimated to be $\sim$2.2$\times$10$^{10}$ cm$^{-3}$ for CRF1 and $\sim$1.9$\times$10$^{10}$ cm$^{-3}$ for CRF2. The peak thermal 
energies of CRF1 and CRF2 are calculated to be 1.81$\times$10$^{30}$ erg and 2.06$\times$10$^{30}$ erg. They are listed in the tenth column of Table~\ref{tab:eng} (see also Figure~\ref{fig12}).
The increases of threshold intensity in 131 {\AA} have marginal influence (5.5\%$-$6.5\%) on the final estimation of peak thermal energies.
Combing the peak thermal energy and radiative loss, the total energy required to heat and maintain the hot plasma at temperatures near 10 MK are $\geq$2.0$\times$10$^{30}$ erg for CRF1
and $\geq$2.2$\times$10$^{30}$ erg for CRF2, respectively.

\subsection{Energy in flare-accelerated electrons} \label{s:nth}
\begin{figure*}
\plotone{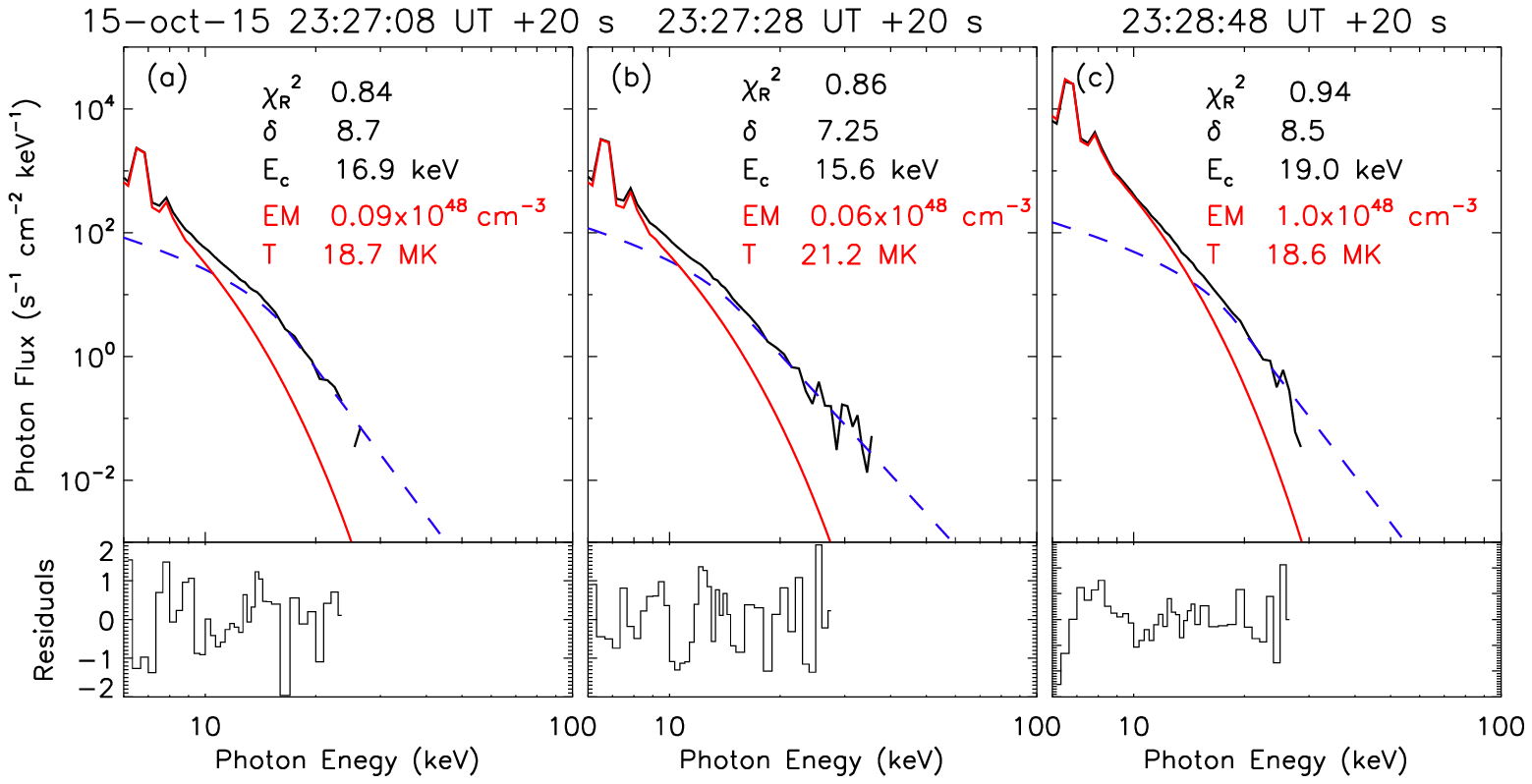}
\caption{Three characteristic HXR spectra (black solid lines) and the fitted thermal (red solid lines) and nonthermal (blue dashed lines) components during CRF1. 
The isothermal temperature and EM of the thermal component are labeled.
The low-energy cutoff ($E_c$), spectral index ($\delta$) of nonthermal electrons, and $\chi_R^2$ of the residuals are also labeled.
\label{fig10}}
\end{figure*}

In Figure~\ref{fig10}, three characteristic \textit{RHESSI} spectra made from detector 5 near the peak time of 25$-$50 keV flux of CRF1 are displayed. The integration time is 20 s \citep{feng13}.

\begin{figure*}
\plotone{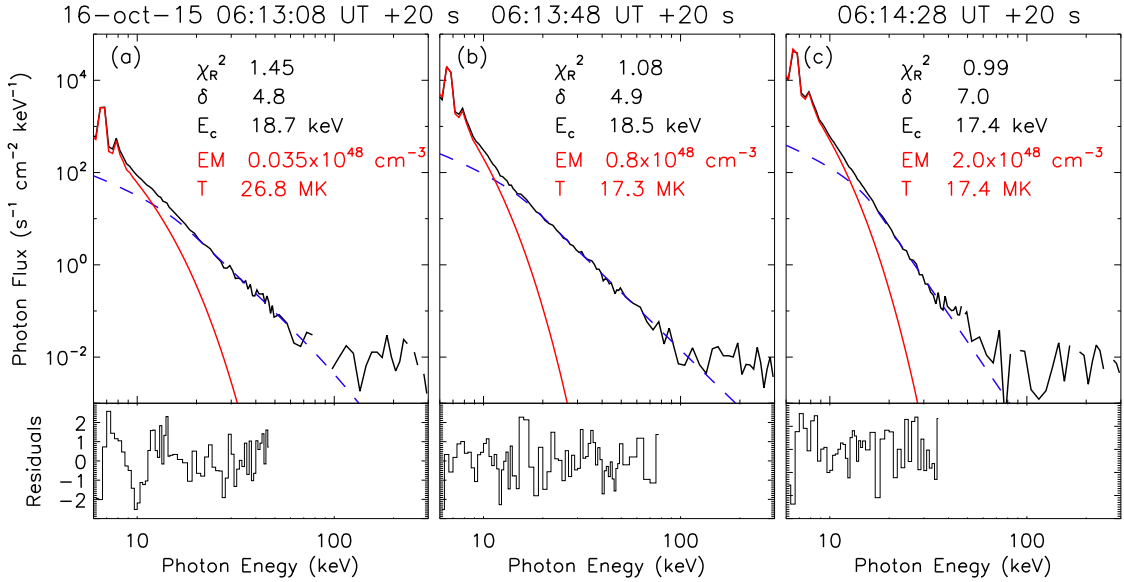}
\caption{Same as Figure~\ref{fig10}, but for CRF2.
\label{fig11}}
\end{figure*}

The distribution of injected nonthermal electrons $F_0(E_0)$ is assumed to be in the form of single power law \citep{sai05}:
\begin{equation} \label{eqn5}
  F_0(E_0)=A_0E_0^{-\delta}, E_0\geq E_{c},
\end{equation}
where $A_0$ is a normalized parameter of the total electron flux (in unit of 10$^{35}$ electrons s$^{-1}$), $E_{c}$ signifies the low-energy cutoff, and $\delta$ denotes the power-law index.
In each panel, the fitted thermal component is drawn with a red line, while the nonthermal component is drawn with a blue line.
Likewise, three characteristic spectra and results of fitting for CRF2 around 06:14 UT are displayed in Figure~\ref{fig11}.
As expected, the isothermal temperature from \textit{RHESSI} observation is systematically higher than that from \textit{GOES} observation, 
which is explained by the fact that \textit{RHESSI} is more sensitive to high-temperature plasma \citep{war16a,war16b}. 
The temperature exceeds 25 MK only at one moment (see Figure~\ref{fig11}(a)), implying the existence of direct-heated super-hot plasma \citep{cas14,war16a,war16b}.
For each flare, EM increases with time, which is consistent with the occurrence of chromospheric evaporation that quickly responds to the injection of electrons \citep{fis85,gra15}.
The values of $\delta$ in the range 4$-$9 are close to those of M1.0 flare on 2010 August 7 \citep{fle13}.

By comparing the spectra in Figure~\ref{fig10} and Figure~\ref{fig11}, it is revealed that the electrons are accelerated to higher energies by CRF2 than CRF1.
Meanwhile, the spectra indices of CRF2 are a factor of $\sim$1.5 harder than those of CRF1, indicating that CRF2 is more energetic than CRF1.
The low-energy cutoffs (15$-$20 keV) of the two flares, however, do not have considerable difference.

The accumulated energy carried by nonthermal electrons above $E_c$ is derived by integrating the injected power with time,
\begin{equation} \label{eqn6}
  E_{nt,e}=A_{nt}\int_{t_1^\prime}^{t_2^\prime} \int_{E_c}^{E_H}E_0F_0(E_0){dE_0dt},
\end{equation}
where $A_{nt}$ is the electron injection area, $E_H$ is fixed at 30 MeV \citep{ems12,war16a}, $t_1^\prime$ and $t_2^\prime$ stand for the lower and upper time limits of integral. 
For CRF1, $t_1^\prime$ and $t_2^\prime$ are set to be 23:27 UT and 23:47 UT. For CRF2, they are set to be 06:11 UT and 06:31 UT, respectively. 
It is noticeable that the predominant input of nonthermal energy happens around the peak time of HXR above 25 keV. The contribution of nonthermal energy well after the peak time is negligible.
The nonthermal energies in electrons are calculated by summing over all time intervals between $t_1^\prime$ and $t_2^\prime$, which are listed in the last column of Table~\ref{tab:eng}.
It is obvious that the nonthermal energy input by flare-accelerated electrons is sufficient to explain the heating requirement of hot plasma including the radiative loss.

\begin{figure}
\plotone{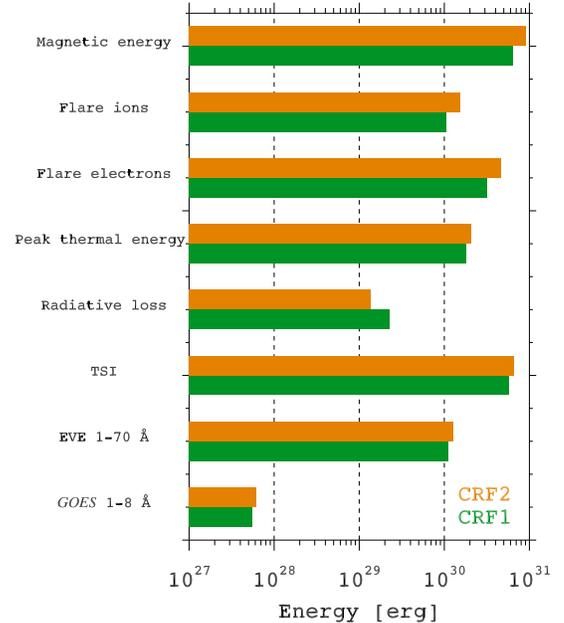}
\caption{Bar chart showing the different energy components of CRF1 (green bars) and CRF2 (orange bars). 
The TSI, nonthermal energy of flare-accelerated ions, and dissipated magnetic energy are estimated from previous statistical results.
\label{fig12}}
\end{figure}

Compared with the energy in electrons, a quantitative calculation of the nonthermal energy in flare-accelerated ions is much more difficult. In this work, we might as well estimate the energy in ions 
according to previous statistical works. The mean ratio of energy in ions to that in electrons is reported to be $\sim$1/3 \citep{ems12,asch17}. Hence, the nonthermal energy in ions for CRF1 and 
CRF2 are 1.1$\times$10$^{30}$ erg and 1.5$\times$10$^{30}$ erg, respectively. The total nonthermal energy including electrons and ions for CRF1 and CRF2 are thus 4.3$\times$10$^{30}$ erg
and 6.1$\times$10$^{30}$ erg (see Figure~\ref{fig12}). Our results are in accordance with previous conclusion that the thermal and nonthermal energies are of the same magnitude \citep{hol03,sai05}.

So far, it is still controversial whether the energy content in flare-accelerated electrons and ions is sufficient to account for the bolometric radiation (i.e., TSI). On one hand, \citet{ems12}
concluded with caution that there is sufficient energy in the flare-accelerated particles to account for the total bolometric radiation, though the mean ratio of energy in accelerated particles to the
bolometric radiated energy is $\sim$0.7 in their sample. On the other hand, \citet{war16b} insisted that the nonthermal energy input by energetic electrons is insufficient to account for the total
heating requirement of the hot plasma or for the bolometric loss, especially for weak flares. In this study, the estimated nonthermal energies of CRFs are very close to the estimated bolometric 
radiation (see Section~\ref{s:xuv170} and Figure~\ref{fig12}). Hence, our result is in favor of the conclusion by \citet{ems12}, albeit in-depth investigation is needed in the future.

In a comprehensive investigation of the global energetics of $\sim$400 flare/CME events observed by \textit{SDO}, \citet{asch17} concluded that nonthermal energy of flare-accelerated particles,
the energy of direct heating, and the energy in CMEs, which are the primary processes in an eruptive flare, account for $\sim$87\% of the dissipated magnetic energy. The nonthermal
energies in electrons and ions account for more than 2/3 of the dissipated energy. Therefore, the dissipated magnetic energies of CRF1 and CRF2 can be estimated, 
being $\sim$6.4$\times$10$^{30}$ erg and $\sim$9.1$\times$10$^{30}$ erg (see Figure~\ref{fig12}). 
It should be emphasized that the estimated dissipated free energy is an upper limit, since the CRFs in our work are confined events without CMEs and a larger part of dissipated energy 
are involved in particle acceleration. The reason why we do not perform a nonlinear force-free field extrapolation is that AR 12434 was close to the limb, so that the measurement of 
photospheric vector magnetograms was less reliable.

\begin{deluxetable*}{ccccccccccc}
\tablecaption{Event list with component energies ($\times$10$^{30}$ erg) and the ratios (CRF2/CRF1) \label{tab:eng}}
\tablecolumns{11}
\tablenum{2}
\tablewidth{0pt}
\tablehead{
\colhead{flare} &
\colhead{Class} &
\colhead{Date} &
\colhead{$t_{sta}$\tablenotemark{a}} &
\colhead{$t_{peak}$\tablenotemark{b}} &
\colhead{$t_{end}$\tablenotemark{c}} &
\colhead{1$-$8 {\AA}\tablenotemark{d}} &
\colhead{1$-$70 {\AA}\tablenotemark{e}} &
\colhead{$T_{rad}$\tablenotemark{f}} &
\colhead{$E_{th}$\tablenotemark{g}} &
\colhead{$E_{nt,e}$\tablenotemark{h}} 
}
\startdata
CRF1 & M1.1 & 15-Oct-15 & 23:27 & 23:31 & 23:50 & 5.50e-3 & 1.12 & 2.26e-1 & 1.81 & 3.2 \\
CRF2 & M1.1 & 16-Oct-15 & 06:11 & 06:16 & 06:35 & 6.24e-3 & 1.28 & 1.41e-1 & 2.06 & 4.6 \\
\hline
Ratio & \nodata & \nodata & \nodata & \nodata & \nodata & 1.13 & 1.14 & 0.62 & 1.14 & 1.44 \\
\enddata
\tablenotetext{\mathrm{a}}{\,\textit{GOES} start time (UT).}
\tablenotetext{\mathrm{b}}{\,\textit{GOES} peak time (UT).}
\tablenotetext{\mathrm{c}}{\,\textit{GOES} end time (UT).}
\tablenotetext{\mathrm{d}}{\,Radiated energy in \textit{GOES} 1$-$8 {\AA}.}
\tablenotetext{\mathrm{e}}{\,Radiated energy in \textit{SDO}/EVE 1$-$70 {\AA}.}
\tablenotetext{\mathrm{f}}{\,Total radiated energy from the SXR-emitting plasma.}
\tablenotetext{\mathrm{g}}{\,Peak thermal energy of the SXR-emitting plasma.}
\tablenotetext{\mathrm{h}}{\,Nonthermal energy in flare-accelerated electrons.}
\end{deluxetable*}

\begin{deluxetable*}{ccccccc}
\tablecaption{Evaluation of the area ($\times$10$^{18}$ cm$^2$) and volume ($\times$10$^{28}$ cm$^3$) of SXR-emitting plasma \label{tab:vol}}
\tablecolumns{7}
\tablenum{3}
\tablewidth{0pt}
\tablehead{
\colhead{flare} &
\colhead{$A_{131}$} &
\colhead{$V_{131}$} &
\colhead{$A_{1600}$} &
\colhead{$V_{1600}$} &
\colhead{$\bar{A}$\tablenotemark{a}} &
\colhead{$\bar{V}$\tablenotemark{b}}
}
\startdata
CRF1 & 6.30 & 1.58 & 5.65 & 1.34 & 5.98 & 1.46 \\
CRF2 & 6.67 & 1.72 & 6.62 & 1.70 & 6.64 & 1.71 \\
\enddata
\tablenotetext{\mathrm{a}}{\,Mean value of $A_{131}$ and $A_{1600}$.}
\tablenotetext{\mathrm{b}}{\,Mean value of $V_{131}$ and $V_{1600}$.}
\end{deluxetable*}

\section{Discussions} \label{s:disc}
As mentioned in Section~\ref{s:intro}, CRFs are confined flares without CMEs in most cases. A few works are dedicated to the energy partition in confined flares until now. 
\citet{th15} studied the homologous confined X-class flares in AR 12192 in 2014 October. The total nonthermal energy in electrons ($\sim$10$^{32}$ erg) of an X1.6 flare, 
accounting for $\sim$10\% of the free magnetic energy stored in the AR, is significantly greater than that in eruptive flares of the same class. 
For the two M1.1 CRFs in this study, the nonthermal energies in electrons ((3.9$\pm$0.7)$\times$10$^{30}$ erg) are comparable to that of M1.2 flare (3.25$\times$10$^{30}$ erg) 
in \citet{war16a} and nearly twice larger than that of M1.2 flare (2.0$\times$10$^{30}$ erg) in \citet{sai05}. Additional statistical study is worthwhile to clarify whether nonthermal energies
in confined M-class flares are substantially larger than in eruptive flares of the same class.

In this work, for the first time we explore the energy partition in two homologous CRFs. There are several factors that may have effect on the estimation of different energy contents.
On one hand, the isothermal temperatures of flares from \textit{GOES} are adopted when calculating the peak thermal energies \citep{ems12,feng13}. 
However, the post-flare loops are multithermal in nature \citep{sun14}. The thermal energies based on DEM analysis are found to be $\sim$14 times on average larger than those of 
isothermal plasma \citep{asch15}. Hence, the peak thermal energies in our study might be underestimated.
On the other hand, the volumes of SXR-emitting plasma in Table~\ref{tab:vol} are almost one order of magnitude larger than the values of M-class flares \citep{sai05}. 
The difference may originate from the different methods of volume calculation. As mentioned above, the threshold intensities of flares in 131 {\AA} also have effect on the volume of thermal plasma.
In this respect, the estimations of thermal energies might be overestimated. 
In brief, the two factors (temperature and volume) play complementary roles, indicating that the results of thermal energies are in a reasonable range. 

The nonthermal energy of electrons is quite sensitive to the low-energy cutoff \citep{asch16}. \citet{sui05} analyzed an M1.2 flare on 2002 April 15 and determined the cutoff energy (24$\pm$2 keV). 
The total nonthermal energy in electrons is calculated to be $\sim$1.6$\times$10$^{30}$ erg. For the two M1.1 flares in our study, the cut-off energies are less than 20 keV (see Figure~\ref{fig10} 
and Figure~\ref{fig11}). In the fourth line of Table~\ref{tab:eng}, the ratios of energy components between CRF2 and CRF1 are listed. It is obvious that CFR2 is somewhat more energetic than CRF1, 
and the energy partition is similar for the homologous flares of the same class.

Finally, it should be emphasized that our calculation of energy partition of CRFs have limitations.
The magnetic free energy, nonthermal energy of flare-accelerated ions, and TSI are estimated based on previous results for the lack of suitable data (see Figure~\ref{fig12}). 
Besides, the conductive loss is not considered, although it might be negligible compared to radiative loss \citep{ems12}.
The energies of CMEs and solar energetic particles accelerated by a CME-driven shock are not considered because of the confined nature of flares.

In the next step, investigation of the energetics of CRFs with an EUV late phase \citep{wood11,dai13} is worthwhile, since part of the thermal energy is contained in the hot spine loops 
during the late phase as long as 1$-$2 hr after the main impulsive phase \citep{sun13}. Furthermore, we will work out the energetics of eruptive CRFs associated with jets \citep{wang12,zqm19b} 
or CMEs \citep{jos15}. The kinetic, potential, and thermal energies distributed in the flare-related jets or CMEs will be evaluated as precise as possible.

\section{Summary} \label{s:sum}
In this paper, we investigate the energy partition of two homologous M1.1 CRFs in AR 12434 on 2015 October 15 and 16. 
The peak thermal energy, nonthermal energy of flare-accelerated electrons, total radiative loss of hot plasma, and radiant energies in 1$-$8 {\AA} and 1$-$70 {\AA} of the flares are calculated.
The main results are summarized below:
\begin{enumerate}
\item{The two flares have similar energetics, with CRF2 being slightly more energetic than CRF1.
The peak thermal energies are (1.81$-$2.06)$\times$10$^{30}$ erg. The nonthermal energies in flare-accelerated electrons are (3.2$-$4.6)$\times$10$^{30}$ erg.
The radiative outputs of the flare loops in 1$-$70 {\AA} ((1.12$-$1.28)$\times$10$^{30}$ erg) are $\sim$200 times greater than the outputs in 1$-$8 {\AA}.}
\item{The radiative losses of SXR-emitting plasma ((1.41$-$2.26)$\times$10$^{29}$ erg) are one order of magnitude lower than the peak thermal energies.
The total heating requirements of flare loops including radiative loss are (2.1$\pm$0.1)$\times$10$^{30}$ erg, which could sufficiently be supplied by nonthermal electrons.}
\item{The dissipated magnetic free energies, nonthermal energies of flare-accelerated ions, and TSI are roughly estimated based on previous statistical results.
In-depth investigations of energy partition of eruptive CRFs and CRFs with EUV late phases will be focused in the future.} 
\end{enumerate}

\begin{acknowledgements}
The authors appreciate the reviewer for valuable suggestions to improve the quality of this paper.
\textit{SDO} is a mission of NASA\rq{}s Living With a Star Program. AIA and HMI data are courtesy of the NASA/\textit{SDO} science teams. 
This work is funded by NSFC grants (No. 11773079, 11790302, 11673048, 11373023, 11961131002, 11573072, 11603077, 11203083, U1731241), 
the Fund of Jiangsu Province (BK20161618, BK20161095), 
the Youth Innovation Promotion Association CAS, the Strategic Pioneer Program on Space Science of CAS (XDA15010600, XDA15052200, XDA15320103, and XDA15320301), 
and the project supported by the Specialized Research Fund for State Key Laboratories. 
L.F. is supported by National key research and development program 2018YFA0404202. 
\end{acknowledgements}

\end{document}